\documentclass[twocolumn,showpacs,preprintnumbers,amsmath,amssymb]{revtex4}
\usepackage{graphicx}
\usepackage{dcolumn}
\usepackage{bm}

\newcommand{\be}{\begin{equation}}
\newcommand{\ee}{\end{equation}}
\newcommand{\ba}{\begin{eqnarray}}
\newcommand{\ea}{\end{eqnarray}}

\begin{document}
\renewcommand{\figurename}{{\bf Fig.}}
\renewcommand{\tablename}{{\bf Tab.}}
\renewcommand{\figurename}{{\bf Fig.}}
\renewcommand{\tablename}{{\bf Tab.}}

\title{${\mathcal N}$ew physics for superluminal particles}
\author{S. D.~Hamieh}
\affiliation{Department of Physics, Lebanese
University, Faculty of Sciences (I), Hadath, Beirut, Lebanon}
\date{\today}
\begin{abstract}
In this paper, we consider the apparent superluminal speed of neutrinos in their
travel from CERN to Gran Susso, as measured by the OPERA experiment, within the framework of
the Extended Lorentz Transformation Model.  The model is
based on a natural extension of Lorentz transformation  by wick rotation.  Scalar and Dirac's fields are considered and  invariance under the new lorentz group is discussed. Moreover,  an
 extension of quantum mechanics to accommodate new particles is considered using  the newly proposed
 Generalized-$\mathbb{C}$ quantum mechanics.
  A two dimensional representation   of the new Dirac's equation
 %, within the framework of GCQM,
 is therefore formulated and its solution is calculated.
\end{abstract}
\maketitle

%Keywords: Dirac's equation, non associative algebra, superluminal Neutrinos.\\

\section{Introduction}
Recently, the OPERA collaboration, according to their precision measurement,   claims \cite{1}  an early arrival time of CNGS
(CERN Neutrino beam to Gran Sasso) muon
antineutrinos traversing 730 kilometers from CERN to Gran Sasso. This corresponds to
${v-c\over c}=2.37 \pm 0.32\times 10^{-5}\,$.
This larger deviation of the neutrino velocity from $c$  is a new result
pointing to new physics in the neutrino sector.
The CNGS neutrinos have an average energy of 17 GeV with a broad distribution reaching up to
several tens of GeV. A separate measurement of neutrinos above and below 20 GeV has revealed no significant
energy-dependence of the superluminality in this energy range.
The OPERA claim is  compatible with earlier results obtained by the MINOS experiment at FERMILAB \cite{2}.
 This result  was recently confirmed in a new investigation by OPERA
  using a beam with a short-bunch  (see table \ref{tab1}). To understand the underlying
physics, a large number of papers has been published in arXiv that can  be categorized
into models of geometric solutions in extra dimensions \cite{4}, deformed special relativity \cite{5} ,
environmental superluminality \cite{6,8}, and explicit Lorentz violation \cite{3}, and
combinations of these ideas. While most of  theories \cite{11,22,33,44}  are concerned about the Lorentz
violation/modification, our main motivation here is the extension of
Lorentz transformation
using a natural mechanism namely a wick rotation via $c\rightarrow ic$.  As consequence of this transformation,
 a new dispersion relation is discovered which  allows to probe a new velocity domain.
The model will be applied to superluminal neutrino to obtain an estimation of neutrino mass. Moreover, as our main concern here is to probe new physics, we have considered the dynamics of superluminal particles not only within the framework of quantum mechanics but also within the framework of generalized quantum mechanics proposed in \cite{9}.

This paper is organized in the following manner. In the next section, the extended Lorentz transformation Model (ELTM), is presented and its application
to neutrino is studied. In section 3 application of ELTM to field theory is discussed.
In section 4 using the discovered dispersion relation a new Dirac's equation (DE) within the framework of
 Generalized-$\mathbb{C}$ quantum mechanics (GCQM) is derived. Section 5 summarizes the results
   of the present investigation and also concluded remarks are given.

\section{Modeling superluminal particles}

The basic idea of the proposed Extended Lorentz Transformation Model (ELTM) is the observation that the Minkowski metric
    $$ds^2 = -(c dt)^2 + dx^2 + dy^2 + dz^2$$ and the four-dimensional Euclidean metric
   $$ds^2 =  (cd\tau)^2 + dx^2 + dy^2 + dz^2$$
are equivalent if we make, Wick rotation,  that is if one permits the coordinate $ct$ to take on imaginary values \footnote{The
 Minkowski metric becomes Euclidean when  $ct$ is restricted to the imaginary axis, and vice versa. Taking a problem expressed in Minkowski space with
 coordinates $x, \,y,\, z,\, t$, and substituting $ct = ic\tau$, sometimes yields a problem in real Euclidean
 coordinates $x, \,y,\, z,\, \tau$, which is easier to solve. This solution may then, under reverse substitution,
  yield a solution to the original problem.}.
Applying this rotation to Lorentz transformation, that is allowing $c$, which is the only parameter of special relativity, to take imaginary value ($c\rightarrow ic$)
  the
space-time transformation becomes
$x'^{\mu}=\Lambda^{\mu}_{\,\,\,\nu}x^{\nu}$
and
$${\Lambda}=\left(
  \begin{array}{cccc}
    \cos\theta & \sin\theta& 0 & 0 \\
    -\sin\theta & \cos\theta & 0 & 0 \\
    0 & 0 & 1 & 0 \\
    0 & 0 & 0 & 1 \\
  \end{array}
\right)\,$$
 with  $\cos \theta=\gamma={1/\sqrt{1+{v^2\over c^2}}}$ and, because of the Euclidian metric, $x^{\mu}=(ct,{\bf x})=x_{\mu}$.
Clearly the symmetry group of such transformation is SO(4) with the properties that
$$Lie SO(4)=Lie SO(3)\bigoplus Lie SO(3)$$
and its  representation is equivalent to the tensor product of the two representations of SU(2) where the objects that transform under this
representation are $$|j_1m_1;j_2m_2\rangle=|j_1m_1\rangle\bigotimes |j_2m_2\rangle$$
%%\ba x'=\gamma (x-vt),\quad y'&=&y\,,\quad z'=z\,,\quad t'=\gamma (t+{v\over c^2}x)\,,\nonumber\\
 %%\quad {\rm with }\quad \gamma&=&{1\over\sqrt{1+v^2/c^2}}\,.\nonumber\ea
We observe that the new transformation extends the Lorentz transformation to  a range of velocity \footnote{The foundations of special relativity are the relativity principle and the invariance
 of the speed of light. As a result, the Galilei group of classical mechanics is
replaced by the Lorentz group, which leads, for example, to the relativistic law of addition
of velocities. The fact that the speed of light is the maximum attainable velocity of all
particles does not directly follow from the Lorentz group, since it only delivers an invariant
velocity at first.}
 $v>c$.
Note that, in the ELTM,  we avoid any necessity for imaginary masses
in order to have $v>c$ as  done in tachyon theory. Thus sidestepping the
 problem of interpreting exactly what a complex-valued mass may
 physically mean.
Other physical quantities can be obtained upon substitution of
 $c\rightarrow ic$, in particular, a new  dispersion relation can be found \footnote{Note that  $p_{\rm max}={m_0c}$  correspond to
$v\rightarrow \infty $ and
$E_{\rm max}={m_0c^2\over\sqrt{2}}$  correspond to
$v=c $.}
\begin{equation} \label{1} E^2=-p^2c^2+m_o^2c^4\,.\end{equation}
\section{Discussion about the physical interpretation of the discovered dispersion relation}
The found dispersion relation (Eq. \ref{1}) can have a physical interpretation as propagation of a wave in a medium with negative index of refraction
 (metamaterial). Moreover, the  mass term $m_o$ can be generated by a quantification in an extradimension without introducing the higgs field \footnote{work in progress}.
In fact,  if we assume  the following equation of motion
$$(\partial_{\mu}\partial^{\mu}+\alpha\partial_5)\phi=0\,,$$
and  by considering
 a wave solution to this equation with
periodic  boundary condition for Bosons \footnote{Note that for fermion we must assume antiperiodic boundary conditions.
}  in the quantified extradimension  we can generate the mass term as
  $$m^2\propto p=\hbar{2\pi n\over a}\,.$$
Here $\alpha$ is  complex and $\partial_5$ is the partial derivative with respect to the extradimension.

  An important observation can be seen from this approach. By evaluating the angular momentum
 $L=\hbar{2\pi n\over a}\times {a\over 2\pi}=\hbar n$ we found that
$$L\propto{a\over 2 \pi }m^2$$
 which is the Regge trajectory, one of the main motivation of string theory.

 Moreover,   the proposed dispersion
relation Eq. 1 introduce a limit to the maximum energy of the superluminal particles thus  no need of renormalization group and
perhaps a unified theory can emerge.

 \begin{table}
\begin{center}
\caption{
The summary of superluminal neutrinos from OPERA, MINOS}
\begin{tabular}{lll}
  \hline
  % after \\: \hline or \cline{col1-col2} \cline{col3-col4} ...
Experiment&Velocity ratios ${v-c\over c}$& Energy range\\
\hline
\hline
OPERA&$(2.37 \pm 0.32) \times 10^{-5}$&17 GeV\\
MINOS&$(5.1 \pm 2.9) \times 10^{-5}$&3 GeV\\
  \hline
\end{tabular}
{\label{tab1}}
\end{center}
\end{table}
\begin{figure}[htb]
\vspace{-3cm}
\centering\includegraphics[width=9cm]{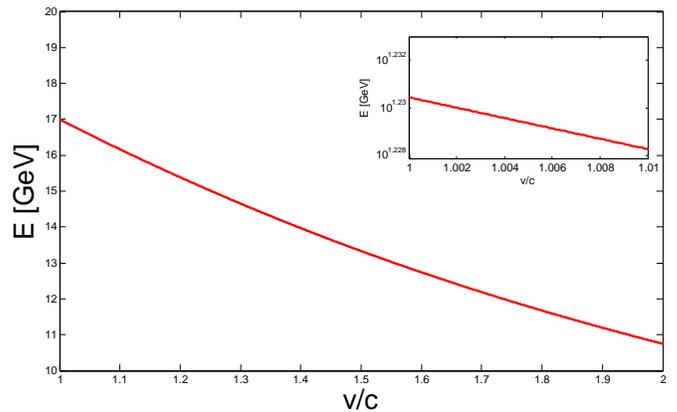}
\vspace{-3.28cm}\caption{$E$ in GeV as function of $v/c$ for superluminal neutrino.
\label{e.pdf}}
\end{figure}
\begin{figure}[htb]
\vspace{-3cm}
\centering\includegraphics[width=9cm]{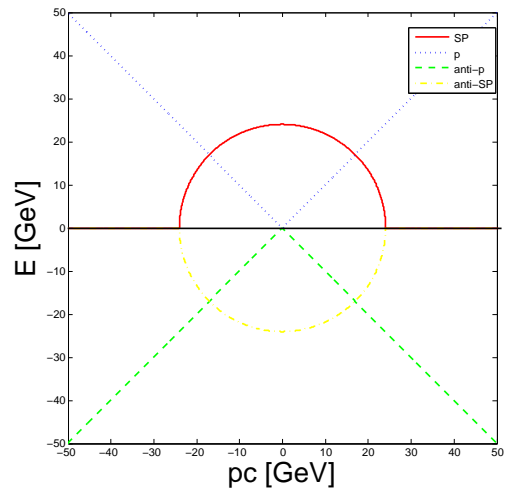}
\vspace{-3.28cm}\caption{$E$ in GeV as function of $pc$ in GeV: Solid line for SP, Dotted line
for particle, Dashed line for antiparticle, Dashed dotted line anti-SP.
\label{dispersion.pdf}}
\end{figure}

\section{Application of the model to the Opera results}

Applying Eq. \ref{1} to the Opera results,   the neutrino mass can be   extracted
\begin{equation} m_{\nu_{\mu}}c^2\sim {\sqrt{2}E}\sim 24.4
\rm GeV \,.\nonumber\end{equation}
 In Fig. \ref{e.pdf} we show  $E$ in GeV versus $v/c$ where we have fixed $E=17$ GeV at $v/c=1$ for neutrino.
 From the inner figure of Fig. 1, we
   conclude that there is no significant
energy-dependence of the superluminality at the vicinity of $v\sim c$. Moreover, this result can be confronted to experimental
results in the future if more superluminal particles (SP) are observed with large  velocity
 ranges. These results, can be subject to two criticisms
\begin{enumerate}
\item
The velocity dependence like shown in  Fig. 1
would lead to much larger deviations of neutrino velocity from $c$ than suggested by Opera: a large range of neutrino
energies from few GeV to 80 GeV needs to be considered.
\item
Also,  Eq. 1 does not allow real energies for neutrino momenta
above $m_0c^2$ = 24.4 GeV. This contradicts the observations from Z decay at LEP where neutrinos with energy 45.6 GeV were produced.
\end{enumerate}

 However, since enough experimental data for mass-matrix of the neutrinos {\bf under consideration} is still absent, it is
 not clear its exact form  for the superluminal neutrino. Here, we claim the following

 {\it ``If  nature allow the existence of superluminal particles  they must obey to the proposed dynamics}."

 In Fig. \ref{dispersion.pdf} we show the neutrino $E$ as function of $pc$ for particle/antiparticle and SP/anti-SP (4 branches). From Fig. \ref{dispersion.pdf} we expect that the neutrino becomes superluminal at the 4 points of  intersection of the
circle describing the Superluminal neutrino and the lines describing neutrino. We call this mechanism as the superluminality mechanism for zero mass particles.
\section{Application of the ELTM to Field theory}
In order to show that our approach  is not  a simple transformation from Minkowski space to Euclidian space,
which transform the  action $S\rightarrow S_E$ \footnote{A transformation from quantum to statistical mechanics },
we consider the application of the model to   complex local  field theory. Consider as an example,
   a scalar field with the  dispersion relation given above (Eq. \ref{1}) and defining
   $$\widehat{\partial}_{\mu}=\left({\partial\over \partial t}\,,\,i{\bf\nabla}\right)\label{mu}\,,$$
    then, the field equation can be written as ($c = \hbar = 1$)
\begin{equation}
(\widehat{\partial}_\mu \widehat{\partial}^\mu+m^2)\phi = 0,\hspace{.3in}\mu=0,1,2,3 \label{2}
\end{equation}
with $E= i\partial_t$ and $p=-i{\bf \nabla} $.
The field equation  written  at this form, that is using the $\widehat{\partial}$,  is used  to conserve
% Minkowski metric and thus regarding
properties like  covariant and contravariant. This will simplify
the discussion of invariance under the following  extended Lorentz transformation
\footnote{Note that, this representation of $\widehat{\Lambda}$ which transform $(ict,{\bf x})\rightarrow (ict',{\bf x}')$ is used to conserve
 the Minkwovsi metric. One can use another variant
of the new Lorentz transformation matrix that transform $(ct,{\bf x})\rightarrow (ct',{\bf x}')$
%which gives  where $\beta'={v\over c}$.
  However, in this case, the invariance is  not straightforward because  the use of Euclidean $g^{\mu\nu}={\rm diag}(1,1,1,1)$ metric is essential.}
$$\widehat{\Lambda}=\left(
  \begin{array}{cccc}
    \gamma & -\beta \gamma& 0 & 0 \\
    -\beta \gamma & \gamma & 0 & 0 \\
    0 & 0 & 1 & 0 \\
    0 & 0 & 0 & 1 \\
  \end{array}
\right)=\left(
  \begin{array}{cccc}
    \cos\theta & i\sin\theta& 0 & 0 \\
    i\sin\theta & \cos\theta & 0 & 0 \\
    0 & 0 & 1 & 0 \\
    0 & 0 & 0 & 1 \\
  \end{array}
\right)$$
where $\beta=-i{v\over c}$, and $\cos\theta=\gamma$.
The invariance of Eq. \ref{2} can be deduced from the invariance of the scalar field  in standard field theory  using the following transformation $\Lambda\rightarrow\widehat{\Lambda}$
and ${\partial}^\mu\rightarrow\widehat{\partial}^\mu$.
%The invariance of Eq. \ref{2} is due to the invariance of the  Euclidean metric tensor $g^{\mu\nu}$ under this transformation.
Now, after studying invariance of the new scalar field, it is useful to see what happen for the dynamical quantities like Green's function. We start from
 the  wick-rotated \footnote{By wick-rotated we mean the transformation $c\rightarrow ic$ }  functional integral of $\phi$
\begin{equation}
\int {\it {\mathcal D}}\phi(x)\exp\left [i\int d^4x{\mathcal L}(\phi)\right ]\nonumber
\end{equation}
with ${\mathcal L}={1\over 2}(\widehat{\partial}_{\mu}\phi)^2-{1\over 2}m^2\phi^2$ is the Lagrangian density for the
free scalar field. The generating function  of $\phi(x)$  is
\begin{equation}
Z(j)=\int {\it {\mathcal D}}\phi(x)\exp\left [i\int d^4x({\mathcal L}(\phi)+j\phi) \right]\nonumber
\end{equation}
Therefore, the Green's function integral of $\phi(x)$  can be
calculated from the generating function  exactly as Minkowski Green's function that is
\ba G_F(x,x')&=&{-1\over Z(0)}{\delta Z\over\delta j(x)\delta j(x')}=
\langle T\phi(x)\phi(x')\rangle \nonumber\\&=&\int {d^4k\over (2\pi)^4} {e^{-ik(x-x')}}{i\over k^2-m^2+i\epsilon}\,.\nonumber\ea
This Green's function,  is just  the Feynman propagator with  $$k^2=w^2-{\bf k^2}\rightarrow k^2=w^2+{\bf k^2}.$$
As for the Dirac field, it is also possible to construct its wave equation.
In fact, the proposed DE for SP that must be invariant under the extended Lorentz transformation is
\[(i\gamma^\mu\widehat{\partial}_\mu - m)\psi = 0\]
Of course, this equation is guessed from the non superluminal particles (NSP) DE. Again, as done for scalar field,
the Lorentz invariance of this equation, under the extended Lorentz group, can be deduced from the  invariance of the
 NSP. In fact, the generators of the Lee algebra of the extended Lorentz group will be the same
 as Lorentz group algebra. The only change is in the antisymmetric tensor that gives the infinitesimal angle where
 $v/c$ must be replaced by $\beta=-iv/c$ (See any standard
 field theory text book). Moreover, solution of this equation can be obtained
from  the NSP solution of DE using the following substitution
\begin{equation}{\bf p}\rightarrow i{\bf p} \label{pp}\end{equation}
in Dirac's spinors.
Thus the positive energy solution four-spinors is
\begin{equation}
u^{(s)}=N
\left(\begin{array}{c} \chi^{(s)} \\ {{\bf i\sigma p}\over m+E } \chi^{(s)}
\end{array}\right) \label{eqn:dm2}\nonumber
\end{equation}
\vspace*{.13in}
and the negative energy solution is
\begin{equation}
u^{(s)}=N
\left(\begin{array}{c} {{\bf -i\sigma p}\over m+|E| } \chi^{(s)}\\ \chi^{(s)}
\end{array}\right) \label{eqn:dm2}\nonumber
\end{equation}
where $N$ is a normalization factor and $\psi=ue^{-ipx}$.
Other physical quantities for Dirac field can be obtained by the same substitution given in Eq. \ref{pp}.
For example when calculating  cross-section for some physical process in standard model  one has to take into account
for SP the $m_{\nu}$ and the substitution of
$$p=(E,{\bf p})\rightarrow p=(E,i{\bf p})\,.$$
For example, the scattering  cross section of superluminal $e^+e^-\rightarrow \mu^+\mu^-$ can be inferred from the subluminal scattering for the same process.
In fact we found
$${\sigma}_{\rm superluminal}={\sigma}_{\rm subluminal}={4\pi\alpha^2\over 3s}$$

Application  of this formalism to interacting fields and calculation of cross section for some physical
process will be left for future investigation.
In next section, as we are interested in new physics that might govern the dynamics of SP,
we take this opportunity to see the smallest  representation of the
new DE in the newly proposed
algebra  \cite{9}. This algebra is called the Generalized-$\mathbb{C} $ (G$\mathbb{C} $)   extend  quantum theory to new class of theories based
on  the non associative algebra.

%%%%%%%%%%%%%%%%%%%%%%%%%%%%%%%%%%%%%%%%%%%%%%%%%%%%%%%%%%%%%%%%%%%%%%%%%%%%%%%%%%%%%%%%%%%%%%%%%%%%%%%%%%%%%%%%%%%%%%
%%%%%%%%%%%%%%%%%%%%%%%%%%%%%%%%%%%%%%%%%%%%%%%%%%%%%%%%%%%%%%%%%%%%%%%%%%%%%%%%%%%%%%%%%%%%%%%%%%%%%%%%%%%%%%%%%%%%%%
%%%%%%%%%%%%%%%%%%%%%%%%%%%%%%%%%%%%%%%%%%%%%%%%%%%%%%%%%%%%%%%%%%%%%%%%%%%%%%%%%%%%%%%%%%%%%%%%%%%%%%%%%%%%%%%%%%%%%%

\section{ Beyond local complex field theory description of SP }
%Two dimensional representation of  the new  DE in the Generalized-$\mathbb{C}$}
%Perhas the dynamics of SP will not follow local complex field theory we will present in this section
%a new form of DE using

Since our aim in this paper is to search for  new physics that describe
 SP then it is of great importance to investigate the new DE within the framework
 of the extended quantum mechanics that has been recently proposed  in
 \cite{9}. In fact,  Quantum mechanics as developed in the standard textbooks, and as applied to
elementary particle physics in the standard model, is understood to be complex
quantum mechanics: The wave functions and probability amplitudes are
represented by complex numbers. However, it has been known since the 1930s that
more general quantum mechanical systems can, in principle, be constructed.

In the
present work, the Generalized-$\mathbb{C} $ quantum mechanics \cite{9} is studied for SP. However, we do not
construct, a general formalism  for Generalized-$\mathbb{C} $ local field theory. We only concentrate on
the description of the new DE within the framework of  Generalized-$\mathbb{C} $ quantum mechanics. We believe that this  will be useful in  future investigation of new physics.
For clarity, to avoid the explicit use of $i$, the most general form  of DE is \footnote{
From now on, the positions of the indices, $\mu, \nu$ etc have no significance with respect to covariance
or contravariance and are placed for typographical convenience.  Repeated indices, however, do indicate summation}
\begin{equation}
\mathcal{H}\psi=(C_\mu\partial_\mu)\psi=(C_x \partial_x + C_y \partial_y + C_z \partial_z + C_t \partial_t)\psi= m\psi. \label{eqn:dw}
\end{equation}
To recover the new Klein-Gordon equation
\begin{equation}
(-\nabla^2 - \partial_t^2)\psi =  m^2\psi , \label{eqn:KG}\nonumber
\end{equation}
the following conditions must hold
\ba C_{x,y,z}^2 = -1;\hspace{.1in}  C_t^2 &=&  -1; \hspace{.1in}\nonumber\\\left\{C_\mu,C_\nu\right\}=C_\mu C_\nu + C_\nu C_\mu &=&0,\nonumber\\ \mbox{ where } \mu\neq\nu.\hspace{.1in}\mu,\nu &=& x,y,z,t\label{eqn:cnd}
\ea
%Equation ($\ref{eqn:dw}$) can be rewritten
%\begin{equation}
%(\gamma^\mu\partial_\mu -m)\psi = 0,\hspace{.3in}\mu=0,1,2,3 \label{eqn:qb}
%\end{equation}
%by defining
%\begin{equation}\gamma^\mu = (C_t,C_x, C_y, C_z)\,,\label{x}\end{equation}
%this avoids the explicit use of an imaginary scalar.
Using the following Dirac matrices,
satisfying ($\ref{eqn:cnd}$)

$$C_t= \left(
\begin{array}{cccc}
0 & e_1 \\
e_1& 0
\end{array} \right)$$
$$C_x  = \left( \begin{array}{cccc}
 0 & e_2 \\
 e_2 & 0
      \end{array} \right)$$
$$C_y= \left(
\begin{array}{cccccccc}
 e_1& 0 \\
0 & -e_1
       \end{array} \right)$$
$$C_z = \left( \begin{array}{cccc}
e_2 &0 \\
0 & -e_2
      \end{array} \right)$$

($e_1, e_2$ are the imaginary G$\mathbb{C}$ units, see Appendix),   in equation($\ref{eqn:dw}$) results in: \vspace*{.2in}
\ba
&&(\mathcal{H}-m) \psi=\nonumber\\&&
\left(\begin{array}{cccc}
-m+e_1\partial_y+e_2\partial_z & e_1\partial_t +e_2\partial_x \\
e_1\partial_t + e_2\partial_x & -m-e_1\partial_y-e_2\partial_z
\end{array}
\right)\left(\begin{array}{c} \psi_1 \\ \psi_2
\end{array}\right)=0 \label{eqn:dm}
\nonumber \ea
\vspace*{.13in}
\newline
The solution to this equation in 1+1 dimension, $y,t$, is
\begin{equation}
\psi(y,t)= N\left(\begin{array}{c} {E\over p+m} \\1
\end{array}\right)e^{e_1(py-Et)}\quad \,  \label{eqn:dm}
\end{equation}
where, as usual  $p$ represents the `momentum', $E$ is the `energy' and $N$ is a normalization factor. As expected  $E=\pm \sqrt{-p^2+m^2}$.
Note that the importance of G$\mathbb{C}$ algebra is in the existence of complex solution in
small dimension (here 2) which is not available in $\mathbb{C}$ algebra.

%%%%%%%%%%%%%%%%%%%%%%%%%%%%%%%%%%%%%%%%%%%%%%Conclusions%%%%%%%%%%%%%%%%%%%%%%%%%%%%%%%%%%%%%%%%%%%%%%%%%%%%%%%%%%%%%%%%%%%%%%%%%%%
%%%%%%%%%%%%%%%%%%%%%%%%%%%%%%%%%%%%%%%%%%%%%%%%%%%%%%%%%%%%%%%%%%%%%%%%%%%%%%%%%%%%%%%%%%%%%%%%%%%%%%%%%%%%%%%%%%%%%%%%%
\section{Conclusions}
We have attempted to account for neutrino superluminality, as reported by OPERA, while staying within the
familiar framework of local field theory. We show that superluminal neutrinos     can be fitted into the original Standard
Model without changing other results.
Since enough experimental data for mass-matrix of the neutrinos is still absent, it is not clear what is the exact form of it.
 However, using the extended Lorentz transformation  model we  have provided an estimation of this mass.
It is also found, in this paper, a two dimensional Dirac's wave function for the new dispersion relation within the framework
of the three dimensional non-associative algebra. Finally  we believe that the  proposed models
merits to be explored in more physical problem.\\

%%%%%%%%%%%%%%%%%%%%%%%%%%%%%%%%%%%%%%%%%%%%%%%%%%%%%%%%%%%%%%%
%%%%%%%%%%%%%%%%%%%%%%%%%%%%%%%%%%%%%%%%%%%%%%%%%%%%%%%%%%%%%
\section*{Appendix}

%%%%%%%%%%%%%%%%%%%%%%%%%%%%%%%%%%%%%%%%%%%%%%%%%%%%%%%%%%%%%%%%%%%%%%%%%%%%%%%%%%%%%%%%%%%%%%%
%But, before considering this case
 We remind the reader about the properties of  the  Generalized-$\mathbb{C}$ algebra (G$\mathbb{C}$).
 An interested reader may refer to \cite{9,10} for further information.
The proposed generalization  of the $\mathbb{C}$ algebra, the
G$\mathbb{C}$, is finite-dimensional
non division algebra  \footnote{ A division algebra,  is a finite
dimensional algebra for which $a\neq0$ and $b\neq 0$ implies
$ab\neq  0$, in other words, which has no nonzero divisors of
zero. }
 containing the real numbers $\mathbb{R}$ as a sub-algebra and  has the following properties:
\begin{itemize}
\item A general G$\mathbb{C}$ number, \emph{q}, can be written as
\[ q = ae_0 + be_1 + ce_2\]
where
\[\emph{a,b,c}
\in \mathbb{R} \,\rm or\, \mathbb{C}\,,\]
The real $e_0=1$ and the imaginary G$\mathbb{C}$ units, $e_1, e_2$ are defined by
\[e_1e_1 = e_2e_2 = -1\,,\quad e_1e_2 = e_2e_1 = 0\]
%Note that every non-zero G$\mathbb{C}$ has an inverse.
\item The addition is defined as
\[q_1 + q_2 = a_1 + a_2 +(b_1 + b_2)e_1 + (c_1 +c_2)e_2\,, \] is associative
\[ q_1 + (q_2 +q_3) = (q_1 + q_2) + q_3\,.\]
\item The multiplication is defined  as
\[q_1q_2 = a_1a_2 - b_1b_2 - c_1c_2  +(a_1b_2 + b_1a_2)e_1 +(a_1c_2 + c_1a_2 )e_2 \,,\]
is non-associative under multiplication that is $(q_1q_2)q_3 \neq q_1(q_2q_3)$.
\item The norm of an element $q$ of  G$\mathbb{C}$ is defined by \[N(q)=(\bar{q}q)^{1/2}=(a^2 + b^2 + c^2)^{1/2}\]
with the   G$\mathbb{C}$ conjugate $\bar{q}$ given by \[\bar{q}=a - be_1 -ce_2\,.\]
\end{itemize}
%After this short introduction about the G$\mathbb{C}$, we are ready to find the  two dimensional representation of the DE. }

\section*{Acknowledgments}
The author is very grateful to D. T. Elkhechen for very useful discussions.
This work was performed as part of the research program of Doctorate School of Science and Technology.

%%%%%%%%%%%%%%%%%%%%%%%%%%%%%%%%%%%%%%%%%%%%%%%%%%%%%%%%%%%%%%%%%%%%%%%%%%%%%%%%%%%%%%%%%%%%%%%%%%%%%%%%%%%%%%%%%%%%%%%%%
%%%%%%%%%%%%%%%%%%%%%%%%%%%%%%%%%%%%%%%%%%%%%%%%%%%%%%%%%%%%%%%%%%%%%%%%%%%%%%%%%%%%%%%%%%%%%%%%%%%%%%%%%%%%%%%%%%%%%%%%%

\end{document}